\documentclass[10pt,journal,compsoc]{IEEEtran}

\ifCLASSOPTIONcompsoc
  \usepackage[nocompress]{cite}
\else
  \usepackage{cite}
\fi

\usepackage{balance}
\usepackage{epstopdf}
\usepackage{graphics}
\usepackage{graphicx}
\usepackage{subfigure}
\usepackage{color}
\usepackage[colorlinks=true, urlcolor=black, linkcolor=black, citecolor=black, pdfborder={0 0 0}]{hyperref}
\usepackage{cite}
\usepackage{amsmath,amssymb,amsfonts}
\usepackage{algorithmic}
\usepackage{graphicx}
\usepackage{textcomp}

\usepackage{xspace}
\usepackage{makecell}
\usepackage{listings}
\usepackage{tikz}
\usepackage{multirow}
\usepackage{longtable}
\usepackage{pgf} 

\newcommand{\ignore}[1]{}

\newcommand\note[1]{{{#1}}}  

\newcounter{packednmbr}

\newenvironment{packeditemize}{
\begin{list}{$\bullet$}{
\setlength{\labelwidth}{8pt}
\setlength{\itemsep}{0pt}
\setlength{\leftmargin}{\labelwidth}
\addtolength{\leftmargin}{\labelsep}
\setlength{\parindent}{0pt}
\setlength{\listparindent}{\parindent}
\setlength{\parsep}{0pt}
\setlength{\topsep}{3pt}}}{\end{list}}

\newcommand{\tabincell}[2]{\begin{tabular}{@{}#1@{}}#2\end{tabular}}

\begin{document}
%
\title{Toward Scalable Fully Homomorphic Encryption Through Light Trusted Computing Assistance}

\author{Wenhao~Wang,~\IEEEmembership{IEEE~Member,}~Yichen~Jiang,~Qintao~Shen,~Weihao~Huang,~Hao~Chen,~Shuang~Wang, \IEEEmembership{IEEE~Senior~Member,} XiaoFeng~Wang,~\IEEEmembership{IEEE~Fellow,} Haixu~Tang, Kai~Chen, Kristin~Lauter and~Dongdai~Lin %
\IEEEcompsocitemizethanks{\IEEEcompsocthanksitem Wenhao Wang, Qintao Shen, Weihao Huang, Kai Chen and Dongdai Lin are with the State Key Laboratory of Information Security, Institute of Information Engineering, Chinese Academy of Sciences, Beijing, China. 
\IEEEcompsocthanksitem Yichen Jiang is with University of Florida at Gainesville, Florida. 
\IEEEcompsocthanksitem Hao Chen and Kristin Lauter are with the Cryptography Group at Microsoft Research in Redmond, Washington.
\IEEEcompsocthanksitem Shuang Wang, XiaoFeng Wang and Haixu Tang are with Indiana University at Bloomington, Indiana.

\IEEEcompsocthanksitem Corresponding Author: Wenhao Wang (wangwenhao@iie.ac.cn).}
}

\IEEEtitleabstractindextext{%
\begin{abstract}
It has been a long standing problem to securely outsource computation tasks to an untrusted party with integrity and confidentiality guarantees. While fully homomorphic encryption (FHE) is a promising technique that allows computations performed on the encrypted data, it suffers from a significant slow down to the computation. In this paper we propose a hybrid solution that uses the latest hardware Trusted Execution Environments (TEEs) to assist FHE by moving the bootstrapping step, which is one of the major obstacles in designing practical FHE schemes, to a secured SGX enclave. TEEFHE, the hybrid system we designed, makes it possible for homomorphic computations to be performed on smaller ciphertext and secret key, providing better performance and lower memory consumption. We make an effort to mitigate side channel leakages within SGX by making the memory access patterns totally independent from the secret information. The evaluation shows that TEEFHE effectively improves the software only FHE schemes in terms of both time and space. 

\end{abstract}

}

\maketitle

\IEEEraisesectionheading{\section{Introduction}}
\label{sec:intro}

\IEEEPARstart{L}ong haunting the security research community and the cloud industry is how to securely outsource data intensive computing tasks to public cloud platforms, such as Amazon AWS, Google Cloud, etc. The demand for such secure computing solutions continues to grow in recent years, with the availability of a huge amount of data that need computing resources to process. A prominent example is genomic data, which are projected to increase at the pace of 2 to 40 exabytes per year~\cite{stephens2015big}. Analyzing such data requires an enormous amount of computing power, even more expected for protecting them from the untrusted computing environments, since the data often contains sensitive information, such as a patient's susceptibility to a certain disease, which should not be exposed to the public cloud without proper agreements in place. For over a decade, this demand has been expected to be addressed by cryptographic innovations, particularly fully homomorphic encryption (FHE) that allows any computation to be executed on encrypted data and returns only encrypted results back to the data owner. However, although impressive progress has been made, the state-of-the-art FHE techniques are long distance away from practical use, incurring a slowdown on the order of $10^6$~\cite{song2018keystone}.



\vspace{3pt}\noindent\textbf{Secure computing supports: limitations}. More specifically, a fully homomorphic encryption scheme supports an unlimited number of addition and multiplication operations, and therefore can theoretically compute any function. In practice, however, most existing schemes use learning-with-error (LWE) or its extension, ring learning-with-error (RLWE), for homomorphic computing, which introduces noise to the ciphertext for each operation; as a result, after a limited number of operations, the ciphertext needs to be ``refreshed'' to reduce the noise to allow the computation to continue.  Otherwise, the encrypted result can no longer be decrypted correctly. Serving this noise removal is a \textit{bootstrapping} step that performs homomorphic evaluation of ciphertext, which is exceedingly expensive, \note{over 6 orders of magnitude slower than the addition operations and 4 orders of magnitude slower than the multiplication operations (Table~\ref{table:benchmark})}.
Leveled homomorphic encryption ({\em a.k.a.} somewhat homomorphic encryption) relaxes the requirement of bootstrapping step in fully homomorphic encryption, however, it can only support a fixed number of accumulated multiplication operations (i.e., the circuit depth). As a result, the encrypted data under leveled homomorphic encryption can only be used for specific tasks with a pre-defined maximum circuit depth and need to be re-encrypted again for supporting other computing tasks. Also, the leveled HE tends to have larger ciphertexts and secret keys, which causes a higher memory consumption and slows down the basic homomorphic addition or multiplication operations when the circuit depth grows.

More recently, a hardware secure computing solution has gained traction. Such an approach utilizes a trusted execution environment (TEE) with isolated memory and computing space, called \textit{enclave}, to process encrypted user data, ensuring that a compromised operating system (OS), virtual machine hypervisor and even the system administrator cannot directly observe the content of the data within the \textit{enclave}. A prominent example is Intel's Software Guard Extensions (SGX)~\cite{mckeen2013innovative}, a feature of Intel's Skylake or higher generation CPUs that enables decryption and then analysis of sensitive user data inside the enclave, to achieve a privacy assurance at scale. With its promise, such TEE-based techniques are known to be vulnerable to \textit{side channel attacks}, in which the OS-level adversary could induce page faults, monitor accessed bits and cache, perform memory access timing checks and others to \textit{infer} the content of protected user data~\cite{xu2015controlled,wang2017leaky}. These threats are known to be application dependent and hard (expensive) to eliminate. Also in terms of performance, it is known that SGX is constrained by a relatively small ($\leq$ 128MB) protected memory and it becomes much slower once the memory use goes beyond the limit~\cite{taassori2018vault}. Further SGX is not presently available on any CPUs that support multi-socket systems. Even though most cloud providers, including Azure, AWS, Google cloud, Alibaba cloud, IBM etc., all have SGX-capable hosts, the deployment of such systems is still limited: particularly, so far only low-end processors are armed with SGX, with their CPU packages including no more than 6 physical cores, not to mention any GPU, or accelerator supports.

\vspace{3pt}\noindent\textbf{TEE-assisted FHE}. With its limited computing capability and less reliable privacy protection, nevertheless we believe the SGX-like TEE offers a new opportunity to enhance the performance of FHE, without undermining its privacy assurance. In our research, we investigated a \textit{hybrid} secure computing model, in which a small number of SGX enclaves are delegated with lightweight cryptographic tasks that are easy to secure and fully within its memory constraint to assist the generic computation on encrypted data outside the enclaves. Our approach, called \textit{TEEFHE}, utilizes the SGX-based TEE to refresh the ciphertext, removing the noise by decrypting the homomorphically encrypted data followed by a re-encryption step. For the sake of simplicity, let us define such a procedure as \textit{SGX bootstrapping} in the rest of this paper. Unlike the original bootstrapping in HE, which reduces the noise level in ciphertext, SGX bootstrapping is able to completely eliminate the noise in the ciphertext. In this way, not only can we reduce the massively intensive bootstrapping step for homomorphic noise reduction, but we can significantly curtail the encryption level to make ciphertext orders of magnitude smaller and computation on it much faster, and ensure the generality of the computation on the encrypted data. The last point is important since all relatively more efficient leveled homomorphic encryption approaches today can only support a limited number of accumulated multiplications based on given application needs. As a result, the ciphertexts they use become application-specific, and could not be reused for a different analysis with deeper circuits. 

In our research, we implemented TEEFHE over SEAL~\cite{laine2018simple}, the HE library developed by the Microsoft Research. Our system has been designed to achieve a high performance of HE computation with the assistance of enclave and also to ensure a high privacy assurance: we reconstructed the encryption and decryption algorithms in the enclave for SGX bootstrapping to make them completely secret-independent and therefore remove the side channel leaks for the step; we also changed the library to avoid more heavyweight SEAL context initialization for each enclave call. Further, to allow concurrently running FHE processes to share a small number of SGX enclaves, TEEFHE includes a scheduler that utilizes the length of a task queue and expected follow-up workload to decide when to bootstrap the computation for each process. We analyzed the security guarantee of TEEFHE and experimentally evaluated its performance, first over a set of benchmark operations (addition, multiplication and bootstrapping) and then over a real-world logistic regression task. The study shows that our approach improves the performance of the bootstrapping step by 4 to 6 orders of magnitude, and multiplication/addition by 2 orders of magnitude. When tested on the HE based logistic regression task for disease prediction in the 2017 iDASH Genome Privacy Competition~\cite{idash2017}, TEEFHE outperformed the SEAL-based, software only implementation by 2 orders, using only 1\% of its memory consumption.

\vspace{3pt}
The contributions of the paper are outlined as follows:

\begin{packeditemize}

\item \textit{New understanding and new techniques}. We made the attempt to combine hardware TEE and cryptographic algorithms for complex and generic secure computation tasks. Our preliminary results demonstrate 
this {\em hybrid} approach can significantly enhance FHE performance (at least 2 orders of magnitude on the logistic regression task) without undermining its security guarantee. We believe that it will lead to more efforts along this promising direction.

\item \textit{Implementation and evaluation}. We designed a hybrid system -- TEEFHE, and evaluated its security 
and performance using both the benchmark operations and a real-world secure machine learning task. The results show that the system is both secure and effective.

\end{packeditemize}

\vspace{3pt}\noindent\textbf{Roadmap}. The rest of the paper is organized as follows: Section~\ref{sec:background} introduces the background for our research; Section~\ref{sec:design} elaborates the design and implementation of TEEFHE, and our analysis of its security guarantee; Section~\ref{sec:eval} describes the evaluation of the performance of TEEFHE; Section~\ref{sec:discuss} discusses the limitations of our current approaches and future research to address them; Section~\ref{sec:related} reviews the related prior research and Section~\ref{sec:conclusion} concludes the paper. 

\section{Background}
\label{sec:background}


\vspace{3pt}\noindent\textbf{(Fully) Homomorphic Encryption.} Homomorphic encryption was first introduced by Rivest, Adleman and Dertouzos in the late 1970's～\cite{rivest1978data}. It is a form of encryption that allows computations to be carried out directly on ciphertext, and the computation result can be retrieved by decrypting the ciphertext, which is the same as if the computations are performed on plaintext. Homomorphic encryption enables the computation to be computed homomorphically on the ciphertext without exposing the secret data to an untrusted party. Hence, it can be applied in a wide range of scenarios, such as secure outsourcing of data and computation (to public/commercial clouds) and secure voting systems, etc. 

It was not until Gentry's seminal work published in 2009 that the first fully homomorphic encryption (FHE) system became possible, allowing for the homomorphic execution of an arbitrary number of both addition and multiplication operations ~\cite{gentry2009fully}. 
An FHE scheme is built upon a somewhat homomorphic encryption (SHE) scheme that can evaluate arithmetic circuits of a limited depth. For all existing homomorphic encryption schemes, a small noise component is added to the ciphertext during encryption to guarantee the security of the scheme.
Computing homomorphically on ciphertexts may accumulate the noise above a certain maximum tolerance, after which the decryption may not give the correct computation results. In order to evaluate arbitrarily complex circuits, a bootstrapping approach is adopted in an FHE scheme to remove the noise after a number of SHE steps, in which the encrypted private key of the SHE scheme is placed inside the public key so that the accurate ciphertext can be obtained by homomorphically evaluating the decryption circuit on the noise-prone ciphertext and the encrypted secret key. 
Notably, the bootstrapping step in FHE is computationally expensive compared with other leveled homomorphic encryption operations, and has been one of the major obstacles to designing practical FHE schemes in real-world applications. 

\vspace{3pt}\noindent\textbf{Implementing Homomorphic Encryption.} 
Gentry's original FHE scheme was first implemented by Gentry and Halevi~\cite{gentry2011implementing}. Although several optimizations have been adopted, the implementation is considered impractical as the size of the ciphertext and the computation time increase drastically with the increase of the security level.
With the development of the second generation FHE schemes based on the Learning With Errors (LWE) problem and its generalization to rings (RLWE), multiple FHE software packages became available, such as HElib~\cite{halevi2015bootstrapping}, HEAAN~\cite{cheon2018bootstrapping}, SEAL~\cite{laine2018simple}, TFHE~\cite{chillotti2016tfhe}, Palisade~\cite{cousins2017palisade} and cuHE~\cite{dai2015cuhe}, etc., which implement various FHE schemes including Brakerski-Gentry-Vaikuntanathan (BGV)~\cite{brakerski2014leveled} and Fan-Vercauteren (FV)~\cite{fan2012somewhat} among others.

In particular, the Simple Encrypted Arithmetic Library (SEAL) is an open source implementation of a variant of the FV scheme, i.e., the Brakerski/Fan-Vercauteren scheme (BFV), and is developed by the Cryptography Research Group at Microsoft Research. It supports the common arithmetic operations over ciphertext, including addition, multiplication and negative, as well as operations between a ciphertext and a plaintext, such as \texttt{AddPlain} and \texttt{MultiplyPlain}. Notably, SEAL possesses a relatively small, standalone code base.


\vspace{3pt}\noindent\textbf{Intel's Software Guard Extension (SGX).}
Intel's Software Guard Extension (SGX) is a recent implementation of the hardware-based Trusted Execution Environment (TEE), which has become widely available in commodity desktop and workstation processors with Skylake/Kabylake micro-architecture. Mainstream cloud service providers including Google, AWS, Azure and Alibaba cloud are planning to provide SGX-enabled instances.

SGX is an x86 instruction extension providing isolated execution environment. The protected area inside the application address space is called an {\em enclave}.
SGX is designed under a strong adversary model, resilient against adversaries with the system privileges or even full control over the physical machine, with only the processor itself as the trusted computing base. 
To provide integrity and confidentiality protection of data running in an enclave, the processor operates SGX codes and data in an encrypted memory region called the {\em processor reserved memory (PRM)}. Extra permission checks are performed by extending the memory controller when the enclave data is accessed. The code and data are only decrypted after they are loaded into the processor caches. The PRM is limited by size ($<128$ MB in the current available hardware), while the computation in PRM induces a small performance overhead (about 10\%).  

The SGX capabilities are encoded as leaf functions of ENCLU/ENCLS instructions. For the ease of development, Intel provides a set of SGX drivers and SDKs for both Windows and Linux operating systems. The switches between the application and the enclave are through {\em ECalls} (Enclave Interface Functions) or {\em OCalls} (Calls outside the Enclave). However, SGX does not support system calls inside the enclave, and as a result, system calls can only be served after the execution mode is switched from the enclave mode to the normal mode.

On the other hand, it was demonstrated that SGX is vulnerable to various side channel attacks ~\cite{xu2015controlled,wang2017leaky,van2017telling,moghimi2017cachezoom,lee2017inferring}. The vast majority of shared resources, such as the the page table, translation look-aside buffer (TLB), branch target buffer and caches can be exploited as side channel leakage sources for the access patterns of enclave executions.

\vspace{3pt}\noindent\textbf{Provisioning Secrets with Remote Attestation.} Besides the above isolated execution protections, SGX enables a user to verify the hardware configuration of a remote platform, ensuring a software entity is running on an Intel SGX-enabled platform protected within an enclave, before to provision the software with secrets and protected data. In the design of SGX, it is achieved by supporting two forms of attestation. 

{\em The local attestation} allows a source enclave to prove its identity and authenticity to a target enclave running on the same platform. When an enclave is loaded and initialized, the enclave's measurement is generated by the trusted processor. During local attestation, a cryptographic report is generated for the source enclave by computing CMAC on the enclave's identity (including the enclave's measurement) using a report key, which can be generated and verified on the same platform.

With the {\em remote attestation}, an enclave can attest to a trusted remote entity, and establish an authenticated communication channel between them. Remote attestation is done with the help of an Intel signed enclave, called the Quote Enclave (QE). QE receives a local attestation report from an enclave, verifies it through a local attestation. It obtains the Provisioning Seal Key to retrieve the Attestation Key and generates an Attestation Signature with the key. Along with the local attestation report, they are passed to an {\em Intel Attestation Service} (IAS) to verify the signature. During remote attestation, a shared key can be established with a key agreement protocol between the attested enclave and the entity acquiring the attestation. After the authenticated communication channel is established, the remote entity can provision the enclave with secrets and protected data.




%


\vspace{3pt}\noindent\textbf{Adversary Model}.  In this paper, we follow the adversary models for both homomorphic encryption and SGX. We consider a semi-honest (honest but curious) adversary, who has system privileges and full control of the operating system, and is willing to perform side channel attacks on the TEE but does not collude with the Intel. Following we summarize what such an adversary can and cannot do:
\begin{packeditemize}
\item Full control over the system, with system privileges and physical access to the platform;
\item Leverage of different side channels, e.g., cache \& page table and branch prediction based side channels to infer the memory access patterns, and instructions with variant latencies, such as the floating point unit (FPU) based side channels;
\item No collusion with Intel.
\end{packeditemize}

Under the above assumptions, the security goal of the system is that the adversary cannot reduce the effort ({\em w.r.t} both time and memory) needed to recover the secret from the equivalent fully homomorphic encryption schemes with software-based bootstrapping implementations.

\ignore{In this paper, we follow the adversary models for both homomorphic encryption and SGX. 
We assume the adversary has system privileges and full control of the system, but is semi-honest (honest but curious) and will not collude with Intel. Furthermore, the adversary can infer the access patterns of the enclave through \note{{\em any}} of the available side channels. In summary, we assume the adversary possesses the following capabilities:
\begin{itemize}
\item Full control over the system, with system privileges and physical access to the platform;
\item Leverage of side channels, e.g., cache \& page table and branch prediction based side channels to infer the memory access patterns;
\item Leverage of side channels due to instructions with variant latencies, such as the floating point unit (FPU) based side channels;
\item No collusion with Intel.
\end{itemize}

With the above assumptions, the security goal of the system is that the adversary cannot reduce the effort (in regard to both time and memory) needed to recover the secret information from the equivalent fully homomorphic encryption schemes with software only bootstrapping implementations.

}
%
%

\section{TEEFHE: Design and Implementation}
\label{sec:design}

In this section, we introduce the design of TEEFHE. Beginning with an overview of the system, we first describe each component, and then explain the implementation based on SGX and SEAL. Our prototype implementation is based upon SEAL version 2.3.
We choose SEAL due to its simple code base and independence from external libraries, which makes it easier, compared with other HE implementations, to port to the secure enclave. \note{We will explore implementing TEEFHE on other FHE schemes in future work.}

\subsection{Overview}
\begin{figure}
{\centering
\includegraphics[width=0.95\columnwidth]{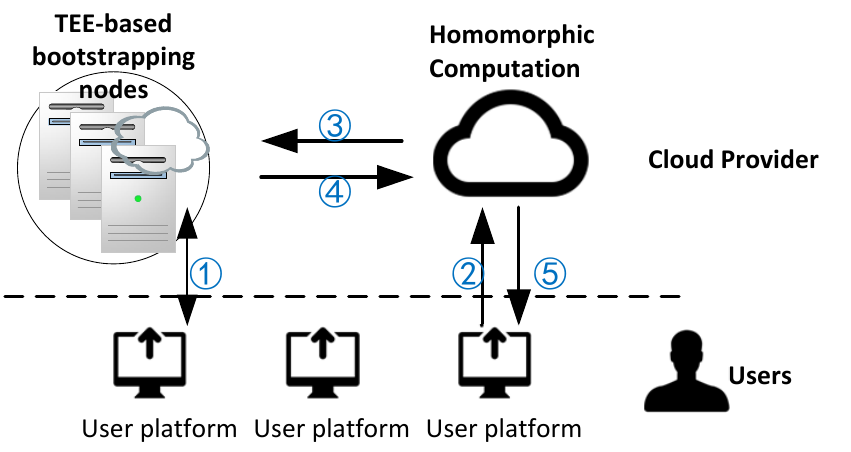}
\caption{Overview of the proposed TEEFHE framework. Participating parties play three different roles, as the users, the homomorphic computation (HC) nodes, or the TEE-based bootstrapping nodes.}
\label{fig:overview}
}
\end{figure}

The structure of TEEFHE is illustrated in Fig.~\ref{fig:overview}. Essentially, the system involves multiple parties playing three key roles in the secure computation, the {\em users} who outsource the computation to a remote party (cloud), with their secret data protected from unauthorized interrogation; the {\em homomorphic computation (HC) nodes} provided by the cloud services using unprotected CPU, GPU or FPGA-based instances; the {\em bootstrapping nodes} provided by the cloud services using the TEE, such as Intel's SGX. The data flow among participating parties with the TEEFHE system is as follows:
\begin{packeditemize}
\item[\textcircled{1}] The users first verify the configuration of the cloud through remote attestation, and establish the shared secret key with the bootstrapping nodes. Afterwards, the users provision their encryption parameters as well as the secret and public keys to the bootstrapping nodes through the established secret channel.
\item[\textcircled{2}] The user's data encrypted under the homomorphic secret key is sent to the HC nodes to perform homomorphic computations. If the computation requires private data from multiple users, each user sends the data encrypted using their own key to the HC nodes.
\item[\textcircled{3}] When bootstrapping is needed in the homomorphic computation, the current intermediate ciphertext is sent from the HC nodes to the bootstrapping nodes. 
\item[\textcircled{4}] The bootstrapping nodes, running inside a secure enclave, first decrypt the ciphertext, then re-encrypt it using the secret key and send the refreshed ciphertext back to the HC nodes. This TEE-based bootstrapping step removes the noise in the ciphertext, and thus enables further homomorphic computation by the HC nodes.
\item[\textcircled{5}] After the whole homomorphic computation is completed, the ciphertext is sent from the HC node back to the users. The users decrypt the ciphertext to retrieve the computation result.
\end{packeditemize}

In this paper, we skip the details of homomorphic computation on the HC nodes. Generally, HC nodes perform the homomorphic operations of the specific somewhat homomorphic encryption (SHE) schemes. We also skip the discussion of the remote attestation, which is a standard procedure within the SGX development framework. 


\subsection{TEE-based Bootstrapping}
A TEE-based bootstrapping primitive accepts a ciphertext as input, decrypts it and re-encrypts the plaintext again to get a refreshed ciphertext. Hence, we implemented the decryption and encryption algorithms within a trusted enclave, which were built to mitigate side channel leakages in the TEE, when the adversary has full control of the platform. On the other hand, to serve multiple bootstrapping requests from the HC nodes (referred to as the ``client'' in the follow-up sections), a scheduling algorithm is developed for the TEE-based bootstrapping nodes (referred to as the ``server'') to utilize the computing and networking resources efficiently.
Below, we present the design of the TEE-based bootstrapping nodes in our prototype implementation of TEEFHE.

\vspace{3pt}\noindent\textbf{Porting SEAL to SGX}. The prototype implementation of TEEFHE provides three interfaces for the application: a \texttt{configure\_para} method to pass the encryption parameters and to set the \texttt{SealContext} object with the given parameters in SGX; a \texttt{set\_key} method to receive the public key and secret key after a remote attestation is performed; and a \texttt{decrease\_noise} method to perform bootstrapping inside SGX, which accepts the ciphertext as the input and returns the refreshed ciphertext.

In our research, we implemented the SEAL functions that are not supported by SGX within the enclave. For instance, as class objects are not allowed as arguments in \texttt{ECall}s or \texttt{OCall}s, we built functions to support saving (or loading) the class objects of \texttt{Ciphertext}, \texttt{Plaintext}, \texttt{PublicKey} or \texttt{Secretkey} to (or from) character buffers. We also modified the random number generator in SEAL so that the hardware sources \texttt{sgx\_rdrand} are always used. 

On the client side, we deployed the \texttt{Simulator} class object of SEAL to estimate the current noise budget after each homomorphic operation. We used the function in SEAL to restore the noise estimation to an initial state after the client's bootstrapping request is served.




\vspace{3pt}\noindent\textbf{Performance enhancement}. In SEALv2.3, a \texttt{SEALContext} object is constructed after all parameters are set. The object class checks the validity and properties of the parameters, and performs several important pre-computations. In its current implementation, a \texttt{SEALContext} object cannot be set as global  because no default constructor is implemented before all parameters are set. As a result, each bootstrapping through the SGX \texttt{ECall} needs to construct a separate \texttt{SEALContext} object, which is more time-consuming (0.1 second over the ring $R = \mathbb{Z}[x]/(x^{4096}+1)$) than the SGX bootstrapping step (0.011 second over the same ring).

In our implementation, we implemented a default constructor for the \texttt{SEALContext} class that allows for global context with uninitialized values, and a \texttt{set\_para($\dots$)} method that allows the \texttt{SEALContext} object to be constructed without an initialization and to be set after the parameters are provided.

\subsection{Scheduling Algorithm}
\begin{figure*}
{\centering
\includegraphics[width=1.98\columnwidth]{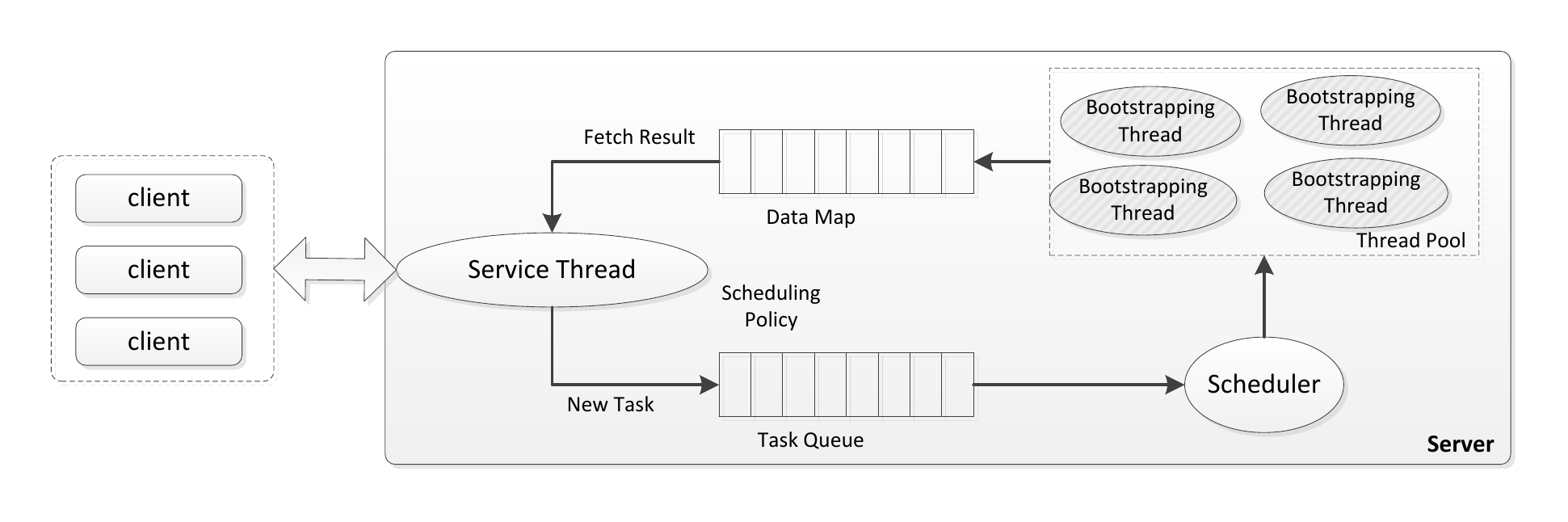}
\vspace{-0.2in}
\caption{The work flow of the scheduling algorithm running on the server (i.e., a TEE-based bootstrapping node). Threads maintained on the server: one service thread, one scheduler thread along with a thread pool consisting of all bootstrapping threads running in an SGX enclave.}
\label{fig:scheduler}
}
\end{figure*}

In the current TEEFHE system, the TEE is used solely for the bootstrapping step, which is achieved through the collaboration between the bootstrapping nodes (the {\em server}) and the HC nodes (the {\em client}).
Notably, one server may accept the bootstrapping requests from multiple clients, whereas one client may send requests to multiple servers, in which all communications are implemented using socket connections. To concurrently serve multiple client requests and make full use of the SGX computing resources, we designed a scheduling algorithm running on the server to manage the execution of these requests, as illustrated in Figure~\ref{fig:scheduler}. 

The server maintains a task queue that stores the bootstrapping tasks, and a data map which stores the ciphertext to be processed. The entries from the task queue and data map are matched through their client ID. A thread pool is created in advance to keep track of the bootstrapping threads inside an SGX enclave. The size of the thread pool is a pre-set parameter that is bounded by the available CPU cores in the enclave.

After each homomorphic operation, each client sends the current noise estimation to the server.
Depending on the message returned from the server, the client may continue the homomorphic computation, or prepare for sending the ciphertext to the server for bootstrapping.

To maintain the communication with the clients, the server will run two system threads, including a service thread responsible of answering the requests from clients, and a scheduler thread responsible of scheduling bootstrapping tasks. The service thread decides whether bootstrapping is needed according to a pre-determined service policy. If it is needed, the service thread inserts a bootstrapping task into the task queue, receives the ciphertext from the client and puts the ciphertext to the data map. In our prototype implementation, we adopted a simple service policy that is solely dependent on the noise budget and the status of the task queue. If the noise budget is above a given threshold of noise level, and the number of the awaiting tasks in the task queue is smaller than 2 times the size of the thread pool, or the noise budget is estimated to exceed the maximum tolerance for decryption, the bootstrapping request is inserted to the task queue and will be served when the computing resource is available.

In the meantime, a scheduler thread on the server fetches a task at a time from the task queue according to the scheduling policy, and launches the bootstrapping thread. In our prototype implementation, we adopted a simple First-Come-First-Served (FCFS) policy for serving the bootstrapping requests. After the ciphertext is refreshed, it is put into the data map with a finished flag set. The service thread periodically queries the status of the data map, and sends the refreshed ciphertext back to the client, until a task's finished flag is set.

\subsection{Side Channel Elimination}
Side channel leakages in SGX have been extensively studied. It has been demonstrated that the enclave execution can be inferred by observing the usage of page tables, branch prediction unit (including branch target buffer and return stack buffer), cache, TLB and DRAM row buffer. In this paper, we categorize the side channel leakages in SGX as follows. \note{To the best of our knowledge, all known side channel attacks (Section~\ref{sec:related}) against SGX fall into these categories.}
\begin{packeditemize}
\item
\textbf{Leakages by secret dependent branches.} The type of side channel leakages associated with leakages from conditional branches, unconditional branches, indirect branches, indirect function calls or function returns. 

For example, the observation of conditional branches reveals the result of a condition evaluation for \texttt{if} or \texttt{loop} statement. A viable method to achieve such observation is the {\em branch shadowing attack}~\cite{lee2017inferring}. As such, the adversary can learn information from the secret if there are secret dependent branches.
\item
\textbf{Leakages by secret dependent memory accesses.}
The type of side channels are related to memory references whose addresses are dependent on the secret information. As such, the adversary capable of collecting memory traces (e.g. by measuring the effect of cache contentions) is able to gain insight into the secrets. 

\item
\textbf{Leakages by instructions with variable execution latency.}  On modern processors, for the sake of performance optimization, a few instructions are implemented in such a manner that their execution latency depends on the operand values. An example is floating-point instructions in x86 platform, because subnormal numbers are rarely encountered, the support for subnormal floating-point numbers is implemented in microcode by the hardware vendors. The execution latency can thus be an order-of-magnitude greater if the operands are subnormal numbers. As such, the adversary with the ability of measuring the instruction latencies may deduce the ranges of the operands.

\end{packeditemize}

\vspace{3pt}
\noindent\textbf{Finding side channel leaks}. To determine side channel leakages, here we first determine the public and secret information, in the sense of whether it is already known to the adversary without looking at a side channel.
\begin{packeditemize}
\item {\em Public information}: public encryption parameters including the coefficient modulus, the polynomial modulus and the plain modulus; 
\item {\em Secret information}: the secret key, the plaintext decrypted from the ciphertext, and any information related such as the length of the plaintext polynomial.
\end{packeditemize}



\begin{table*}
\centering
{\footnotesize
\caption{A summary of modified code segments to remove secret dependent access patterns.}
\vspace{-0.1in}
\label{table:modified}
\begin{tabular}{m{1.7in}<{\centering}|m{1.7in}<{\centering}|m{2.4in}<{\centering}}
\Xhline{1pt}
File & Function & Changes \\ \Xhline{0.5pt}
util/uintarithsmallmod.cpp; util/uintarithmod.cpp & exponentiate\_uint\_mod& rewrite \texttt{if} statement with \texttt{cmov} instructions \\ \hline
encryptor.cpp; rnsencryptor.cpp & Encryptor::preencrypt; RNSEncryptor::rns\_preencrypt &  rewrite \texttt{if} statement with \texttt{cmov} instructions \\ \hline 
decryptor.cpp; rnsdecryptor.cpp & Decryptor::decrypt; RNSDecryptor::rns\_decrypt &  rewrite \texttt{if} statement with \texttt{cmov} instructions \\ \hline
smallntt.h & inverse\_ntt\_negacyclic\_harvey & rewrite \texttt{if} statement with \texttt{cmov} instructions \\ \hline
encryptor.cpp & Encryptor::preencrypt &  resize the destination to size \texttt{coeff\_count} \\ \hline
\Xhline{1pt}
\end{tabular}
}
\end{table*}


The first type of side channel leakages that we try to identify are all branches and memory accesses that have dependency on the secret key or the plaintext. First we used customized Pintools~\cite{luk2005pin} to analyze the basic-block level execution traces and memory traces. Pin is a dynamic binary instrumentation framework that enables the creation of dynamic program analysis tools. As done in previous research~\cite{ohrimenko2016oblivious}, we used Pin on the simulation enclave, which has the same memory layout as the hardware enclave. 

To find side channel leaks, we performed a differential analysis, in which we use Pintool to monitor the changes in basic-block level execution traces and memory traces in the presence of identical and then different secrets (secret keys and plaintext). Particularly, to avoid the noise introduced by randomness during the encryption and decryption operations, we temporarily fixed the outputs of pseudorandom number generator (RNG), as well as the primitive root, so the same precomputed number theoretical transform (NTT) tables will be built when the input secrets are the same. Then we analyze the execution traces to identify the program locations where the same traces are observed given the same input secrets, and different ones are found when the inputs change. Such locations are considered to be secret-dependent. The same technique has also been applied to find secret-dependent memory addresses. After that, we manually locate the functions including these instructions and memory operations using \path{objdump}, and identify their corresponding source codes, which are secret dependent branches and memory accesses.

\ignore{The first type of side channel leakages that we try to identify are all branches and memory accesses that have dependency on the secret key or the plaintext. First we used customized Pintools~\cite{luk2005pin} to analyze the basic-block level execution traces and memory traces. Pin is a dynamic binary instrumentation framework that enables the creation of dynamic program analysis tools. As done in previous research~\cite{ohrimenko2016oblivious}, we used Pin on the simulation enclave, which has the same memory layout as the hardware enclave.
Using randomly generated public keys, secret keys and plaintexts, we collected a set of basic-block level execution traces and memory traces\footnote{When traces were collected, we temporarily used a fixed pseudorandom number generator and a fixed primitive root to generate the precomputed number theoretical transform (NTT) tables, which have no dependency on the secret information.}. Then we identified the differences in the traces, along with the instruction and memory addresses that lead to the differences, by simply comparing them. After that we looked into the source code associated with the (instruction or memory) addresses and located the secret dependent branches and memory accesses.
}
All the secret dependent code segments we found are listed in Table~\ref{table:modified}. One demonstrating example is that SEAL adopts the square-and-multiply algorithm to compute the modular exponentiation of big integer numbers (in file \path{util/uintarithsmallmod.cpp}, see Fig.~\ref{lst:squaremul} for details). The algorithm is known to leak information about the \texttt{exponent} through the observation whether the conditional branch is taken (line 10 to 13), which can be exploited, for example, by a branch shadowing attack~\cite{lee2017inferring}.



The second type of side channels we identified are the instructions with variable latencies. According to the Intel 64 and IA-32 Architectures Optimization Reference Manual~\cite[Appendix C.3, LATENCY AND THROUGHPUT]{inteloptimization}, the instructions with variable latencies depending on the operand values are listed in Table~\ref{table:latency}. We located the references of these instructions in the decryption and encryption procedure by manually inspecting their assembly code with the help of \path{objdump} and confirmed that all of them  are independent from the secret key and the plaintext: they either use constants or random numbers as inputs or read from easily verifiable public sources. As an example, the instruction \texttt{SQRTSD} is used in the function \path{Encryptor::set_poly_coeffs_normal} to generate a noise value sampled from a clipped normal distribution and does not rely on secret information.




\begin{table}
\centering
{\footnotesize
\caption{Instructions with variant latencies of Skylake micro-architecture.}
\vspace{-0.1in}
\label{table:latency}
\begin{tabular}{m{0.95in}<{\centering}|m{1.65in}<{\centering}|m{0.5in}<{\centering}}
\Xhline{1pt}
Instruction & Appeared in function  & \tabincell{c}{Secret\\dependent?} \\ \Xhline{1pt}
SQRTPD; SQRTSD; SQRTPS; SQRTSS & Encryptor::set\_poly\_coeffs\_normal  & no \\ \hline
DIV; IDIV; DIVPD; DIVSD; DIVPS; DIVSS & Encryptor::Encryptor; SealContext::validate & no \\ \hline
VPMASKMOVD/Q & not found & - \\ \hline
RDRAND & sgx\_rdrand & no \\ \hline
CLFLUSH; CLFLUSHOPT & not found & -\\ \hline
\Xhline{1pt}
\end{tabular}
}
\end{table}









\vspace{3pt}
\noindent\textbf{Removing side channel leaks}.
We removed the side channel leakages related to the \texttt{if} condition (shown in Table~\ref{table:modified}) by rewriting the code with conditional move instructions. For example for the code segment shown in Figure~\ref{lst:squaremul}, the value of \texttt{product} is conditionally moved to \texttt{intermediate} depending on whether the condition \path{exponent & 1} is true.
The code for the \path{conditional_mov} function is shown in Fig.~\ref{lst:condmov}. The function takes the source address and destination address as inputs. Depending on whether the condition is satisfied, the consecutive memory data of the given size will be moved from the source address to the destination address (or not). 
The \path{conditional_mov} function is implemented with \texttt{CMOVcc} instructions, which are independent from branch prediction and do not introduce measurable micro-architectural effects that depend on the condition~\cite{Coppens2009Practical}.

The last one of the side channel leakages shown in Table~\ref{table:modified} is that the numbers of coefficients in the plaintext polynomial are leaked during the execution of function \texttt{Encryptor::preencrypt}. The leakage is caused by a \texttt{for} loop, in which the terminating condition depends on the numbers of coefficients in the plaintext polynomial. To remove the leakage, the size of the plaintext is extended to $n+1$ (which is public information) with \texttt{resize} function after it is decrypted (by modifying the \texttt{Decryptor::decrypt} function in the file \texttt{decrypt.cpp}). We further removed the secret dependent \texttt{if} branches in the \texttt{resize} function.


\begin{figure}
\centering
\begin{lstlisting}
uint64_t exponentiate_uint_mod(uint64_t operand, uint64_t exponent, const SmallModulus &modulus) {
@\textcolor{black}-@    ... // fast cases

     uint64_t power = operand;
     uint64_t product = 0;
     uint64_t intermediate = 1;            
    // Initially: power = operand and intermediate = 1, product is irrelevant.
    
     while (true) {
@\textcolor{black}-\textcolor{black}{~~~~~~~~~if (exponent \& 1) \{ } @
@\textcolor{black}-~~~~~~~~~~~~~\textcolor{black}{product = multiply\_uint\_uint\_mod(power, intermediate, modulus);}@
@\textcolor{black}-~~~~~~~~~~~~~\textcolor{black}{swap(product, intermediate);}@
@\textcolor{black}-~~~~~~~~~\textcolor{black}\}@
@\textcolor{black}{+~~~~~~~~~product = multiply\_uint\_uint\_mod(power, intermediate, modulus);}@
@\textcolor{black}{+~~~~~~~~~conditional\_mov(\&product, \&intermediate, 1, exponent \& 1);}@ 
                                 
          exponent >>= 1;
          if (exponent == 0) {
              break;
          }
          product = multiply_uint_uint_mod(power, power, modulus);
          swap(product, power);
      }
      return intermediate;
}
\end{lstlisting}
\caption{The square-and-multiply implementation to compute the modular exponentiation of big integer numbers in SEALv2.3. The \texttt{if} statement (lines beginning with $-$) leaks information of the \texttt{exponent} through side channels, and is replaced with code using conditional move instructions (lines beginning with $+$).}\label{lst:squaremul}
\end{figure}

\begin{figure}
\centering
\begin{lstlisting}
// size: number of qwords (8 bytes)
void __attribute__ ((noinline)) conditional_mov(void *source, void *dest, uint64_t size, uint64_t cond) {
    __asm__ __volatile__ (
        "movq %1, %%rax\n"
        "movq %2, %%rbx\n"
        "movq $0, %%rdx\n"
        "loop:\n"
        "cmpq %3, %%rdx\n"
        "jge exit\n"
        "movq (%%rbx), %%rcx\n"
        "cmp $1, %0\n"
        "cmove (%%rax), %%rcx\n"
        "movq %%rcx, (%%rbx)\n"
        "addq $8, %%rax\n"
        "addq $8, %%rbx\n" 
        "inc %%rdx\n"
        "jmp loop\n"
        "exit:\n" 
        :: "r" (cond), "r" (source), "r" (dest), "r" (size)
        : "rax", "rbx", "rcx", "rdx", "memory");
}
\end{lstlisting}
\caption{The implementation of the \texttt{conditional\_mov} function. We confirmed that the size of data to be moved does not leak private information.}\label{lst:condmov}
\end{figure}




\vspace{3pt}\noindent\textbf{Security analysis}.
With the help of Pin, we confirmed by experiment that the basic block level traces and memory traces during SGX bootstrapping are independent from the secret information. 
As far as we can tell, the secret leakages of the instructions with variant latencies have also been removed. As such we conclude that:
{the adversary's knowledge of the secret information is not extended by observing side channel leakages in the SGX bootstrapping.}

As the HC nodes perform conventional somewhat homomorphic operations, the observations are already available to the adversary. We further examine the network traffics between the bootstrapping nodes and HC nodes, including:
\begin{packeditemize}
\item The size of the transferred data. It is equal to the size of a ciphertext, and can be computed from the public information.
\item Estimation of the noise budget. It is public information obtained from the homomorphic operations that have been performed.
\item The scheduling status and function calls of \texttt{ECall}s and \texttt{OCall}s. They do not reveal information about the HE scheme or the secret key and plaintext.
\end{packeditemize}
Furthermore, with the assumption that the adversary does not collude with Intel, the adversary cannot directly access or interfere with the code and data within the enclaves; otherwise, it would violate SGX's confidentiality and integrity guarantees.

In conclusion, the adversary is not able to gain additional secret information from TEEFHE, and TEEFHE provides data protection of the same security level as the software-based implementation of the underlying FHE scheme.







\section{Performance Evaluation}
\label{sec:eval}
 
Our testbed is equipped with an Intel Xeon E3-1280 v5 processor at 3.7 GHz with 64 GB memory and Hyper-Threading enabled, running a Ubuntu 16.04.2 system with kernel version 4.4.0. The code is compiled with g++-5.4.0. We use the SGX SDK version 2.2. The code of the prototype TEEFHE system consists of 1570 lines of C/C++ codes, in which 703 lines of code are written to port SEAL v2.3 into SGX, and to eliminate the side channel leakages, 386 lines for the socket communication, and 481 lines for the scheduling algorithm. In this section, we evaluate our TEEFHE implementation by addressing the following questions:
\begin{packeditemize}
\item[(1)] How is the performance of SGX bootstrapping?
\item[(2)] How effective is TEEFHE on a real computing task compared with SEAL itself?
\item[(3)] How does the scheduling algorithm scale when handling multiple client requests?
\end{packeditemize}

We follow the notations in SEAL v2.3 as shown in Table~\ref{table:parameters}. 
The choice of encryption parameters significantly affects the performance, capabilities, and security of the encryption scheme. Using a greater
$n$ allows for a greater $q$ to be used without decreasing the security level, and thus the depth of the homomorphic operations becomes greater. On the other hand, a greater $n$ will also decrease the performance (Table~\ref{table:benchmark}).

\begin{table}
\centering
{\footnotesize
\caption{Notations used for the parameters.}
\label{table:parameters}
\begin{tabular}{m{0.45in}<{\centering}|m{1.55in}<{\centering}|m{0.9in}<{\centering}}
\Xhline{1pt}
Parameter & Description & Name in SEAL \\ \Xhline{0.7pt}
$n$ & A power of $2$ & - \\ \hline
$x^n+1$ & Polynomial modulus which defines a ring $R = \mathbb{R}/(x^n+1)$ & \texttt{poly\_modulus}\\ \hline
$q$ & Modulus in the ciphertext space of the form $q_1 \times q_2  \times\dots \times q_k$ & \texttt{coef\_modulus}\\ \hline
$t$ & Modulus in the plaintext space & \texttt{plain\_modulus} \\ \hline
$\chi$ & Error distribution & \\ \hline
$\delta$ & Standard deviation of $\chi$ & \path{noise_standard_deviation} \\ \hline
$B$ & Bound on the distribution $\chi$ & \path{noise_max_deviation} \\
\Xhline{1pt}
\end{tabular}
}
\end{table}

\ignore{
\subsection{Benchmarking TEEFHE}
\begin{table}
\centering
{\footnotesize
\caption{Benchmarking basic operations under different $n$'s and the default $q$'s providing 128-bit security (unit: micro-second). The time is measured with tasks running on a single core.
The dash symbol indicates that a software-based bootstrapping cannot be performed for the given parameters.
}
\label{table:benchmark}
\begin{tabular}{m{0.7in}<{\centering}|m{0.35in}<{\centering}|m{0.35in}<{\centering}|m{0.5in}<{\centering}|m{0.5in}<{\centering}}
\Xhline{1pt}
 $n$ &  4096 & 8192 & 16384 & 32768\\ \Xhline{1pt}
 encryption &  4208  & 9725 & 25211 & 70950\\ \hline
 decryption & 510 & 1897 & 7422& 28142\\ \hline
 addition & 14 & 58& 242& 923\\ \hline
 multiplication & 5221 & 19974 & 83991 & 357404\\ \hline
 square &  3505 & 13533& 57933& 251999\\ \hline
 relinearization &  711& 3956& 25042& 160077\\  \hline
 software-only bootstrapping & - & - & $1.09\times 10^9$ & $2.46*10^{10}$\\  \hline
 SGX bootstrapping & 7881 & 25044 & 87840 & $1.00*10^6$\\
\Xhline{1pt}
\end{tabular}
}
\end{table}
}

\subsection{Benchmarking TEEFHE}
\begin{table}
\centering
{\footnotesize
\caption{Benchmarking basic operations under different $n$'s and the default $q$'s providing 80-bit security (unit: micro-second). The time is measured with tasks running on a single core.
The dash symbol indicates that a software only bootstrapping cannot be performed for the given parameters.
}
\label{table:benchmark}
\begin{tabular}{m{0.7in}<{\centering}|m{0.25in}<{\centering}|m{0.25in}<{\centering}|m{0.25in}<{\centering}|m{0.45in}<{\centering}|m{0.45in}<{\centering}}
\Xhline{1pt}
 $n$ & 2048 &  4096 & 8192 & 16384 & 32768\\ \Xhline{1pt}
 encryption & 2323 & 4704  & 12809 & 36322 & 119422\\ \hline
 decryption & 419 & 878 & 3802 & 13431& 50933\\ \hline
 addition & 17 & 62 & 330 & 1126& 4258\\ \hline
 multiplication & 3839 & 7886 & 38088 & 154495 & 694098\\ \hline
 square & 2818 & 5743 & 28571 & 113490& 522905\\ \hline
 relinearization & 465 & 989& 9012 & 51738& 348991\\  \hline
 software only bootstrapping & - & - & - & {$2.09\times 10^9$} & $5.20\times 10^{10}$\\  \hline
 SGX bootstrapping & 5454 & 11254 & 43750 & $1.43\times 10^5$ &  $1.44\times10^6$\\
\Xhline{1pt}
\end{tabular}
}
\end{table}

We ran a few benchmarks with different parameters, in which $n$ varies from 2048 to 32768, while $q$ is selected accordingly to ensure 80-bit security. The results are summarized in Table~\ref{table:benchmark}. 
It can be seen that when $n$ becomes larger, the running time of all operations increases significantly. For example, the time for homomorphic evaluation of a multiplication increases for $180.8\times$. 
We also observe the increased memory usage for larger $n$. 

As expected, the performance of SGX bootstrapping is dependent on the encryption and decryption time, which is on the same order of magnitude as the homomorphic multiplication, a bit slower than a simple encryption and decryption when $n$ is small, since extra time is spent on transferring the ciphertext across enclave boundaries. When $n$ is large, however, the SGX-based bootstrapping becomes slower, which can be caused by cache misses or page faults as it now has a larger memory footprint.


Since software only bootstrapping is not currently supported in the latest SEAL version 2.3.1, we requested a development version of SEAL from the Microsoft Research.
Compared with the implemented bootstrapping in the development version of SEAL, the SGX bootstrapping has a performance gain of over 4 orders of magnitude when the same parameters are used. For example, when $n = 16384$ the time for the software only and SGX bootstrapping is 2089 seconds and 0.088 second respectively. 

\vspace{3pt}\noindent\textbf{Parameter selection}. 
From Table~\ref{table:benchmark} it can be seen that bootstrapping cannot be supported for smaller $n$'s (i.e. $n = 2048, 4096, 8192$) in SEAL, since the bootstrapping circuit already exceeds the maximum depth that can be correctly evaluated by the SHE scheme.

With the assistance of TEE-based bootstrapping, however, an SHE scheme can be bootstrapped to an FHE scheme as long as it supports at least one multiplication and/or one addition. In our experiment we found that the SHE scheme is bootstrappable even for $n = 2048$ (with proper $q$ providing 80-bit security). Since the TEE-based bootstrapping is fast, it can be more efficient for small $n$ though such a parameter causes more bootstrapping. 
When smaller $n$ is used, the other homomorphic operations such as addition and multiplication, are also much faster in comparison with the setting of $n=32768$. It also incurs less memory usage when smaller $n$ is in use and the computation is performed over a smaller ring. 

\vspace{3pt}\noindent\textbf{Overhead of removing side channels}. To understand the overhead induced by removing side channels, we benchmarked SGX bootstrapping (with {\em vs.} without side channels), averaged over 1000 measurements, and found no observable performance difference. It is reasonable since we only added a bit extra computation to remove the side channels.

\subsection{Evaluating TEEFHE on a Logistic Regression Task}

In this section we report our evaluation on the TEEFHE-based implementation of logistic regression against the software only implementation, adopting the data set and the security requirements of the 2017 iDASH Genome Privacy Competition~\cite{idash2017}. 

\vspace{3pt}\noindent\textbf{Logistic Regression Task.} 
In the 2017 iDASH Genomic Privacy Competition, the participating teams were given genotype/phenotype data about two cohorts (disease vs. healthy), and were challenged to develop a logistic regression model to predict the disease. The computations were required to be performed on ciphertext in order to protect the sensitive health data. The testing data set contains 500 records and 5 binary covariates. The competition requires the solutions provide 80-bit security.


\vspace{3pt}\noindent\textbf{Evaluation Setting.} Parameters were set as follows. 
\begin{packeditemize}
\item\textit{Software only implementation with SEAL.} We adopted the parameters with $n = 32768$ and $q$ of bit length $1020$, which offers 80-bit security\footnote{Note that a large $n$ is necessary for the software only implementation to support bootstrapping and to increase the depth before bootstrapping is needed.}. At most 10 iterations can be performed before bootstrapping is needed.

\item\textit{TEEFHE.} We used two parameter settings for TEEFHE: $n=4096$ (and $n=2048$) with $q$ providing 80-bit security, with which 5 (and 505) bootstrapping are needed for each iteration respectively.

\end{packeditemize} 

We used a single thread to perform the homomorphic computations and  bootstrapping for both the software-based and SGX bootstrapping.



\begin{table}
\centering
{\footnotesize
\caption{Evaluation on a logistic regression task (time measured in seconds). Parameters are set to ensure 80-bit security level.}
\vspace{-0.1in}
\label{table:lreval}
\begin{tabular}{m{0.65in}<{\centering}|m{0.51in}<{\centering}|m{0.56in}<{\centering}|m{0.37in}<{\centering}|m{0.54in}<{\centering}}
\Xhline{1pt}
& \tabincell{c}{Key\\Generation} &  \tabincell{c}{Data\\Preparation}  & Iteration & Memory \\ \Xhline{0.7pt}
\tabincell{c}{Software Only\\(n = 32768)} & 13.71  & 447.11 & 6697.24  & 50.02 GiB \\ \hline
TEEFHE (n=4096) & 0.19  & 15.36  & 80.83 &  455.78 MiB \\ \hline
TEEFHE (n=2048) & 0.09  & 7.60  & 35.41 &  225.86 MiB \\
\Xhline{1pt}
\end{tabular}
}
\end{table}

\vspace{3pt}\noindent\textbf{Evaluation Result.}
The experiment results are summarized in Table~\ref{table:lreval}. 
Since smaller $n$ can be used with TEEFHE, compared with software only implementation,
the running time for the key generation and data encryption is reduced significantly, e.g. the time for key generation was reduced from 13.71 seconds to 0.19 (and 0.09) second.
The memory usage has also dropped, from 50.02 GiB for the software-based solution to 455.78 (and 225.86) MiB for the TEEFHE implementation. Overall, the computation in TEEFHE is about 2 orders faster and the memory usage is less than 1\% of the software only implementation. 

\subsection{Evaluating the Bootstrapping Scheduler}

\begin{figure}
{\centering
\includegraphics[height=0.6\columnwidth,width=0.95\columnwidth]{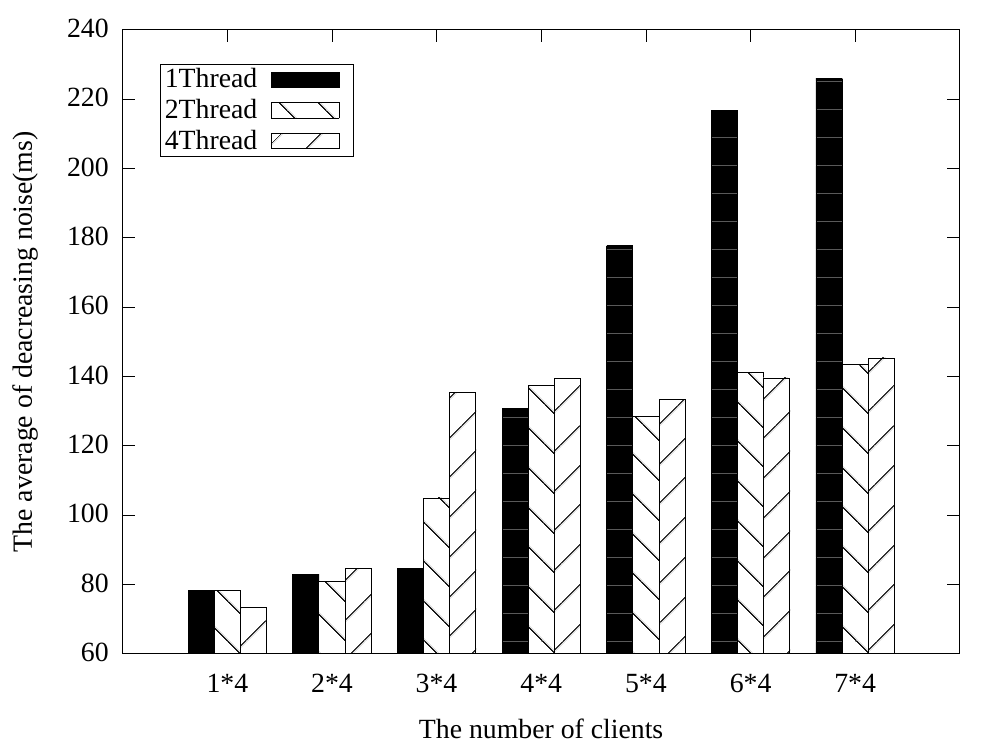}
\vspace{-0.1in}
\caption{Average waiting time for each client bootstrapping request, with the clients number ranging from 4 to 28, and server threads number ranging from 1 to 4.}
\label{fig:schedulereval}
}
\end{figure}

We emulated 4 to 28 clients requesting bootstrapping services from 4 machines to a single SGX-enabled bootstrapping server with a thread pool of a size between 1 and 4. All the client and server machines are equipped with an Intel Xeon E3-1280 v5 processor at 3.7 GHz with 64 GB memory. The clients and the server are connected in a local network environment with 118 MB/s bandwidth and latency of 0.0259 ms.

In our experiment, the parameters for FHE were set as $n = 8192$ and $q$ of bit length 219. Each client ran homomorphic operations in a busy loop of 500 iterations. For each iteration, the client performed a homomorphic addition followed by a homomorphic multiplication. Then the client estimated the current noise budget using the \texttt{Simulator} class object. Afterwards, a request for bootstrapping along with the noise estimation was sent to the server. 
We evaluated the average waiting time (the time between a bootstrapping request is inserted to the task queue and the server returns the refreshed ciphertext) for each thread's request. The result is shown in Figure~\ref{fig:schedulereval}. 
It can be seen that the average waiting time begins to increase when the client number is greater than 8;
when the server has SGX thread pool of size 1, the average waiting time can increase up to over 200 ms, indicating the SGX thread can no longer handle the clients' bootstrapping requests in time, and as a result, the noise level in the client exceeds the threshold, and must wait for bootstrapping task in the task queue to be completed. 
When the size of the SGX thread pool on the server increases to 2 or 4, the scheduler will decide whether the bootstrapping request can be served earlier depending on the current length of the task queue. As a result, the waiting time for the request is never too long, showing that the scheduler works well balancing the workload. 

A notable observation from the result, however, is that when there are 12 clients, the waiting time is even longer if the server has a thread pool of size 4. One possible explanation is that the client bootstrapping requests are served more aggressively, which causes the performance degradation due to the contention of CPU and EPC memory resources, because the processor only has 4 physical cores and less than 128 MiB available EPC memory. We will further explore the issue in future work.

\subsection{Summary}
Overall, for all evaluated parameters the SGX bootstrapping is efficient and over 4 orders of magnitude faster than the software only bootstrapping when performed over the same ring. Furthermore TEEFHE enables the use of smaller $n$'s as long as 1 homomorphic addition and multiplication can be evaluated with the underlying SHE scheme. Using a smaller $n=2048$ the performance gain over $n=32768$ is over $180 \times$ for homomorphic operations, and 6 orders of magnitude for bootstrapping (5454 micro-seconds {\em vs.} 52031 seconds). The memory consumption is also brought down since the computation is performed over a smaller ring. While running homomorphic evaluation on a logistic regression task, the overall speed up is of 2 orders and the memory consumption is brought down to less than 1\%.




\section{Discussion and Future Works}
\label{sec:discuss}

In this paper we proposed TEEFHE, a hybrid solution combining hardware TEEs and homomorphic encryption schemes. We have implemented a prototype TEEFHE based on Intel's SGX and SEAL, and evaluated it on a logistic regression task. The results showed that TEEFHE exhibits a promising improvement over software only implementation of FHE schemes in terms of performance. 

Nevertheless, we believe the following research directions are needed to further explore the capability of bridging the gap between hardware TEEs and conventional cryptographic schemes to gain better performance and to push the cryptographic schemes towards practical use.
\begin{packeditemize}
\item Applying TEEFHE to more application scenarios. Besides logistic regression, it deserves more research efforts to explore the applications of TEEFHE on other privacy preserving computation tasks (including machine learning algorithms) on biomedical data, especially on genome data, etc., in which privacy and data protection are of great importance.
\item Supporting more homomorphic encryption schemes. It is beneficial to realize TEEFHE supporting other SHE schemes, specifically those GPU-based implementations such as cuHE, to further accelerate the homomorphic operations. Another possible future work is exploring the possibility of bootstrapping those SHE schemes that have previously been regarded as non-bootstrappable.
\item Importing more computations into the TEE. Currently, we only replaced bootstrapping within SGX. Because it only involves decryption and re-encryption, the side channel risks are relatively easier to understand and mitigate. Importing more time-consuming computations into the TEE could significantly boost the performance, but may also introduce more side channel attack surfaces, including not only the side channels within the computations in the enclave, but also the network traffic and the interface function calls (\texttt{ECall}s and \texttt{OCall}s). 
\end{packeditemize}

\section{Related works}
\label{sec:related}

\vspace{3pt}
\noindent\textbf{Optimizations and implementations of homomorphic encryption schemes.}
After the first fully homomorphic encryption scheme was introduced by Gentry in STOC'2009~\cite{gentry2009fully}, the optimizations and implementations of homomorphic encryptions schemes have been drawing more and more attentions of researchers.
Gentry's original work presents a general framework of constructing homomorphic schemes, however the secret key needs to be encrypted and made public for the decryption circuit to use. The first generation of the FHE schemes focuses on minimizing the size of secret key and the ciphertext~\cite{smart2010fully,stehle2010faster,gentry2011implementing,gentry2011fully,gentry2012better,gentry2012fully}.

Homomorphic encryption schemes with the security based on the hardness of learning with errors (LWE) problem were presented by Brakerski and Vaikuntanathan since 2011~\cite{brakerski2011fully,brakerski2014efficient}, which marks the beginning of the second generation of homomorphic encryption schemes. Their works introduced a relinearization technique to obtain a somewhat homomorphic encryption that does not require hardness assumptions on ideals, and a dimension-modulus reduction technique to shorten the ciphertexts and reduce the decryption complexity. Later in 2014, the Brakerski-Gentry-Vaikuntanathan cryptosystem (BGV) scheme was proposed to construct leveled FHE schemes which are capable of evaluating arbitrary polynomial-size circuits of a priori bounded depth~\cite{brakerski2014leveled}. Other optimizations include Brakerski's scale-invariant scheme~\cite{brakerski2012fully}, the Gentry-Sahai-Waters scheme (GSW)~\cite{gentry2013homomorphic}, and the Fan-Vercauteren cryptosystem (FV)~\cite{fan2012somewhat}, etc.


The first reported implementation of fully homomorphic encryption is the Gentry-Halevi implementation of Gentry's original scheme~\cite{gentry2011implementing}. Now a few open source implementations of the second-generation FHE scheme are reported~\cite{halevi2015bootstrapping,laine2018simple,cheon2018bootstrapping,dai2015cuhe,al2018implementation,cousins2017palisade,al2018high,ducas2015fhew,chillotti2016faster,chillotti2016tfhe}. These are implemented on general computing resources such as CPU and GPU, and do not utilize the recent TEE techniques.



\vspace{3pt}
\noindent\textbf{SGX side channels.}
Although the design of SGX prevents direct access to the enclave code and data, it has been demonstrated that information leakage from the enclave is possible through many kinds of side channels.

The first demonstrated type of side channel attacks is the controlled channel attacks~\cite{xu2015controlled}, which infer the page level memory access patterns of an enclave by setting and resetting the present bit of the page table entries (PTEs). 
Controlled channel attacks induce a huge number of page faults and 
can be detected within the enclave with the help of Transactional Synchronization Extensions (TSX)~\cite{shih2017t,chen2017detecting}.
However the variant of the controlled channel attack can still work by observing the accessed flags in the PTEs~\cite{wang2017leaky, van2017telling}. 
There are also side channel attacks targeting other competitive use of resources, such as the caches~\cite{gotzfried2017cache,moghimi2017cachezoom,hahnel2017high}, branch target buffer~\cite{lee2017inferring}, translation look-aside buffer~\cite{gras2018translation}, store buffer~\cite{moghimi2018memjam}, cache directories~\cite{yan2019attack}, MMU~\cite{van2018malicious} and DRAM row buffer~\cite{pessl2016drama}. Side channels related to instructions with variant timing, such as \texttt{rdseed}~\cite{evtyushkin2016covert} and floating point instructions~\cite{andrysco2015subnormal} are also published. 
The side channel threat against SGX can be more dangerous since the attacker can precisely control the enclave execution with SGX-Step~\cite{van2017sgx}.

Recently, side channel attacks named meltdown~\cite{lipp2018meltdown} and spectre~\cite{kocher2018spectre} exploiting the out-of-order execution engine of modern processors have gained much attention. The variants can also be applied to SGX~\cite{chen2018sgxpectre,bulck2018foreshadow}. 
The vulnerabilities have been fixed by the recent microcode update
which can be verified remotely by checking the CPU security version numbers (CPUSVN) through remote attestation.


\vspace{3pt}
\noindent\textbf{SGX secured systems and applications.} The use of SGX in many scenarios has been studied, e.g., supporting secure distributed data analytics in the cloud~\cite{schuster2015vc3,zheng2017opaque}, secure networking~\cite{duan2017lightbox,kim2017enhancing,coughlin2017trusted}, privacy preserving biomedical analysis~\cite{chen2016premix,chen2016princess, chen2017presage}, etc. It has also been developed to support secure database queries~\cite{priebe2018enclavedb,eskandarian2017oblivious,gribov2017stealthdb,eskandarian2017oblidb}. 

More related to this work are a series of researches using SGX to enhance cryptographic applications, e.g., supporting secure two party~\cite{gupta2016using} and multi-party computation~\cite{kuccuk2016exploring}. Iron~\cite{fisch2017iron} utilizes SGX to construct secure and practical functional encryption primitives. The core of Iron is a key management enclave to generate encryption keys and signing keys and authorize functions upon a function request. When the function is recovered in a decryption enclave, the function code can be executed in the decryption enclave or the function enclave. Iron makes uses of SGX's attestation service and designs provable secure protocols to ensure the security of the system. In the design of Iron, the function codes are executed within the enclave; while in our TEEFHE design, the operations are executed homomorphically outside the enclave.

The idea of combining SGX and homomorphic encryption are proposed in very recent works~\cite{jiang2018securelr,sadat2018secure,sadat2018safety,chenghong2017scotch}. In these works, only certain operations are performed homomorphically outside the enclave; Otherwise, the ciphertexts are sent to the enclave for decryption and further computation. While the idea is similar, our work exhibits significant difference in the following aspects: \textcircled{1} considering the side channel threat within SGX we imported a minimum computation into SGX and made effort to make it side channel resilient; \textcircled{2} TEEFHE supports arbitrary computation to be performed without pre-configuring the enclave. \textcircled{3} Our design also introduced a schedule process to fully utilize the limited TEE resources to accelerate the whole computation process. We argue that these are of great importance in the design of a hybrid system.

\section{Conclusion}
\label{sec:conclusion}

In this paper we proposed a hybrid solution combining hardware trusted execution environment (TEE) and homomorphic encryption schemes. The proposed TEEFHE system achieves both security and efficiency in a way that replaces the time consuming bootstrapping with a decryption-and-re-encryption step in an SGX enclave. All side channel leakages within the enclave are carefully removed. TEEFHE not only offers a speed-up to the bootstrapping operation, but also enables the somewhat homomorphic encryption to be bootstrappable with small secret key and ciphertext (and thus accelerates basic homomorphic operations such as addition and multiplication) and reduces the memory consumption in the homomorphic computation. Consistent with this expectation, the evaluation of our TEEFHE implementation showed that it achieved a significant performance improvement in terms of both performance and memory consumption.



\bibliographystyle{IEEEtran}
\bibliography{ref}

\begin{thebibliography}{10}
\providecommand{\url}[1]{#1}
\csname url@samestyle\endcsname
\providecommand{\newblock}{\relax}
\providecommand{\bibinfo}[2]{#2}
\providecommand{\BIBentrySTDinterwordspacing}{\spaceskip=0pt\relax}
\providecommand{\BIBentryALTinterwordstretchfactor}{4}
\providecommand{\BIBentryALTinterwordspacing}{\spaceskip=\fontdimen2\font plus
\BIBentryALTinterwordstretchfactor\fontdimen3\font minus
  \fontdimen4\font\relax}
\providecommand{\BIBforeignlanguage}[2]{{%
\expandafter\ifx\csname l@#1\endcsname\relax
\typeout{** WARNING: IEEEtran.bst: No hyphenation pattern has been}%
\typeout{** loaded for the language `#1'. Using the pattern for}%
\typeout{** the default language instead.}%
\else
\language=\csname l@#1\endcsname
\fi
#2}}
\providecommand{\BIBdecl}{\relax}
\BIBdecl

\bibitem{stephens2015big}
Z.~D. Stephens, S.~Y. Lee, F.~Faghri, R.~H. Campbell, C.~Zhai, M.~J. Efron,
  R.~Iyer, M.~C. Schatz, S.~Sinha, and G.~E. Robinson, ``Big data: astronomical
  or genomical?'' \emph{PLoS Biol}, vol.~13, no.~7, p. e1002195, 2015.

\bibitem{song2018keystone}
``Towards an open-source, formally-verified secure enclave,''
  \url{https://keystone-enclave.org/files/dawn-nsf-2018-v5.pdf}, accessed Aug.
  5, 2018.

\bibitem{mckeen2013innovative}
F.~McKeen, I.~Alexandrovich, A.~Berenzon, C.~V. Rozas, H.~Shafi, V.~Shanbhogue,
  and U.~R. Savagaonkar, ``Innovative instructions and software model for
  isolated execution.'' \emph{HASP@ ISCA}, vol.~10, 2013.

\bibitem{xu2015controlled}
Y.~Xu, W.~Cui, and M.~Peinado, ``Controlled-channel attacks: Deterministic side
  channels for untrusted operating systems,'' in \emph{IEEE Symposium on
  Security and Privacy (SP)}.\hskip 1em plus 0.5em minus 0.4em\relax IEEE,
  2015, pp. 640--656.

\bibitem{wang2017leaky}
W.~Wang, G.~Chen, X.~Pan, Y.~Zhang, X.~Wang, V.~Bindschaedler, H.~Tang, and
  C.~A. Gunter, ``Leaky cauldron on the dark land: Understanding memory
  side-channel hazards in sgx,'' in \emph{Proceedings of the 2017 ACM SIGSAC
  Conference on Computer and Communications Security}.\hskip 1em plus 0.5em
  minus 0.4em\relax ACM, 2017, pp. 2421--2434.

\bibitem{taassori2018vault}
M.~Taassori, A.~Shafiee, and R.~Balasubramonian, ``Vault: Reducing paging
  overheads in sgx with efficient integrity verification structures,'' in
  \emph{Proceedings of the Twenty-Third International Conference on
  Architectural Support for Programming Languages and Operating Systems}.\hskip
  1em plus 0.5em minus 0.4em\relax ACM, 2018, pp. 665--678.

\bibitem{laine2018simple}
K.~Laine, ``Simple encrypted arithmetic library-seal (v2.3.1),'' Technical
  report, Tech. Rep., 2018.

\bibitem{idash2017}
``Idash privacy \& security workshop 2017,''
  \url{http://www.humangenomeprivacy.org/2017/}, accessed Aug. 5, 2018.

\bibitem{rivest1978data}
R.~L. Rivest, L.~Adleman, and M.~L. Dertouzos, ``On data banks and privacy
  homomorphisms,'' \emph{Foundations of secure computation}, vol.~4, no.~11,
  pp. 169--180, 1978.

\bibitem{gentry2009fully}
C.~Gentry, ``Fully homomorphic encryption using ideal lattices,'' in
  \emph{Proceedings of the 41st annual ACM symposium on Symposium on theory of
  computing-STOC'09}.\hskip 1em plus 0.5em minus 0.4em\relax ACM Press, 2009,
  pp. 169--169.

\bibitem{gentry2011implementing}
C.~Gentry and S.~Halevi, ``Implementing gentry’s fully-homomorphic encryption
  scheme,'' in \emph{Annual international conference on the theory and
  applications of cryptographic techniques}.\hskip 1em plus 0.5em minus
  0.4em\relax Springer, 2011, pp. 129--148.

\bibitem{halevi2015bootstrapping}
S.~Halevi and V.~Shoup, ``Bootstrapping for helib,'' in \emph{Annual
  International conference on the theory and applications of cryptographic
  techniques}.\hskip 1em plus 0.5em minus 0.4em\relax Springer, 2015, pp.
  641--670.

\bibitem{cheon2018bootstrapping}
J.~H. Cheon, K.~Han, A.~Kim, M.~Kim, and Y.~Song, ``Bootstrapping for
  approximate homomorphic encryption,'' in \emph{Annual International
  Conference on the Theory and Applications of Cryptographic Techniques}.\hskip
  1em plus 0.5em minus 0.4em\relax Springer, 2018, pp. 360--384.

\bibitem{chillotti2016tfhe}
I.~Chillotti, N.~Gama, M.~Georgieva, and M.~Izabach{\`e}ne, ``Tfhe: Fast fully
  homomorphic encryption library over the torus,'' 2016.

\bibitem{cousins2017palisade}
D.~Cousins, Y.~Polyakov, K.~Rohloff, G.~Ryan, G.~Sahu, H.~Sajjadpour, and
  E.~Savas, ``The palisade lattice crypto library,'' 2017.

\bibitem{dai2015cuhe}
W.~Dai and B.~Sunar, ``cuhe: A homomorphic encryption accelerator library,'' in
  \emph{International Conference on Cryptography and Information Security in
  the Balkans}.\hskip 1em plus 0.5em minus 0.4em\relax Springer, 2015, pp.
  169--186.

\bibitem{brakerski2014leveled}
Z.~Brakerski, C.~Gentry, and V.~Vaikuntanathan, ``(leveled) fully homomorphic
  encryption without bootstrapping,'' \emph{ACM Transactions on Computation
  Theory (TOCT)}, vol.~6, no.~3, p.~13, 2014.

\bibitem{fan2012somewhat}
J.~Fan and F.~Vercauteren, ``Somewhat practical fully homomorphic encryption.''
  \emph{IACR Cryptology ePrint Archive}, vol. 2012, p. 144, 2012.

\bibitem{van2017telling}
J.~Van~Bulck, N.~Weichbrodt, R.~Kapitza, F.~Piessens, and R.~Strackx, ``Telling
  your secrets without page faults: Stealthy page table-based attacks on
  enclaved execution,'' in \emph{Proceedings of the 26th USENIX Security
  Symposium. USENIX Association}, 2017.

\bibitem{moghimi2017cachezoom}
A.~Moghimi, G.~Irazoqui, and T.~Eisenbarth, ``Cachezoom: How sgx amplifies the
  power of cache attacks,'' in \emph{International Conference on Cryptographic
  Hardware and Embedded Systems}.\hskip 1em plus 0.5em minus 0.4em\relax
  Springer, 2017, pp. 69--90.

\bibitem{lee2017inferring}
S.~Lee, M.-W. Shih, P.~Gera, T.~Kim, H.~Kim, and M.~Peinado, ``Inferring
  fine-grained control flow inside sgx enclaves with branch shadowing,'' in
  \emph{26th USENIX Security Symposium, USENIX Security}, 2017, pp. 16--18.

\bibitem{luk2005pin}
C.-K. Luk, R.~Cohn, R.~Muth, H.~Patil, A.~Klauser, G.~Lowney, S.~Wallace, V.~J.
  Reddi, and K.~Hazelwood, ``Pin: building customized program analysis tools
  with dynamic instrumentation,'' in \emph{Acm sigplan notices}, vol.~40,
  no.~6.\hskip 1em plus 0.5em minus 0.4em\relax ACM, 2005, pp. 190--200.

\bibitem{ohrimenko2016oblivious}
O.~Ohrimenko, F.~Schuster, C.~Fournet, A.~Mehta, S.~Nowozin, K.~Vaswani, and
  M.~Costa, ``Oblivious multi-party machine learning on trusted processors.''
  in \emph{USENIX Security Symposium}, 2016, pp. 619--636.

\bibitem{inteloptimization}
``{Intel} 64 and {IA}-32 architectures optimization reference manual,''
  \url{https://software.intel.com/sites/default/files/managed/9e/bc/64-ia-32-architectures-optimization-manual.pdf},
  2018, order Number: 248966-040, April 2018.

\bibitem{Coppens2009Practical}
B.~Coppens, I.~Verbauwhede, K.~D. Bosschere, and B.~D. Sutter, ``Practical
  mitigations for timing-based side-channel attacks on modern x86 processors,''
  in \emph{IEEE Symposium on Security and Privacy (SP)}, 2009.

\bibitem{smart2010fully}
N.~P. Smart and F.~Vercauteren, ``Fully homomorphic encryption with relatively
  small key and ciphertext sizes,'' in \emph{International Workshop on Public
  Key Cryptography}.\hskip 1em plus 0.5em minus 0.4em\relax Springer, 2010, pp.
  420--443.

\bibitem{stehle2010faster}
D.~Stehl{\'e} and R.~Steinfeld, ``Faster fully homomorphic encryption,'' in
  \emph{International Conference on the Theory and Application of Cryptology
  and Information Security}.\hskip 1em plus 0.5em minus 0.4em\relax Springer,
  2010, pp. 377--394.

\bibitem{gentry2011fully}
C.~Gentry and S.~Halevi, ``Fully homomorphic encryption without squashing using
  depth-3 arithmetic circuits,'' in \emph{Foundations of Computer Science
  (FOCS), 2011 IEEE 52nd Annual Symposium on}.\hskip 1em plus 0.5em minus
  0.4em\relax IEEE, 2011, pp. 107--109.

\bibitem{gentry2012better}
C.~Gentry, S.~Halevi, and N.~P. Smart, ``Better bootstrapping in fully
  homomorphic encryption,'' in \emph{International Workshop on Public Key
  Cryptography}.\hskip 1em plus 0.5em minus 0.4em\relax Springer, 2012, pp.
  1--16.

\bibitem{gentry2012fully}
------, ``Fully homomorphic encryption with polylog overhead,'' in \emph{Annual
  International Conference on the Theory and Applications of Cryptographic
  Techniques}.\hskip 1em plus 0.5em minus 0.4em\relax Springer, 2012, pp.
  465--482.

\bibitem{brakerski2011fully}
Z.~Brakerski and V.~Vaikuntanathan, ``Fully homomorphic encryption from
  ring-lwe and security for key dependent messages,'' in \emph{Annual
  cryptology conference}.\hskip 1em plus 0.5em minus 0.4em\relax Springer,
  2011, pp. 505--524.

\bibitem{brakerski2014efficient}
------, ``Efficient fully homomorphic encryption from (standard) lwe,''
  \emph{SIAM Journal on Computing}, vol.~43, no.~2, pp. 831--871, 2014.

\bibitem{brakerski2012fully}
Z.~Brakerski, ``Fully homomorphic encryption without modulus switching from
  classical gapsvp,'' in \emph{Advances in cryptology--crypto 2012}.\hskip 1em
  plus 0.5em minus 0.4em\relax Springer, 2012, pp. 868--886.

\bibitem{gentry2013homomorphic}
C.~Gentry, A.~Sahai, and B.~Waters, ``Homomorphic encryption from learning with
  errors: Conceptually-simpler, asymptotically-faster, attribute-based,'' in
  \emph{Advances in Cryptology--CRYPTO 2013}.\hskip 1em plus 0.5em minus
  0.4em\relax Springer, 2013, pp. 75--92.

\bibitem{al2018implementation}
A.~Al~Badawi, Y.~Polyakov, K.~M.~M. Aung, B.~Veeravalli, and K.~Rohloff,
  ``Implementation and performance evaluation of rns variants of the bfv
  homomorphic encryption scheme.'' \emph{IACR Cryptology ePrint Archive}, vol.
  2018, p. 589, 2018.

\bibitem{al2018high}
A.~Al~Badawi, B.~Veeravalli, C.~F. Mun, and K.~M.~M. Aung, ``High-performance
  fv somewhat homomorphic encryption on gpus: An implementation using cuda,''
  \emph{IACR Transactions on Cryptographic Hardware and Embedded Systems}, vol.
  2018, no.~2, pp. 70--95, 2018.

\bibitem{ducas2015fhew}
L.~Ducas and D.~Micciancio, ``Fhew: bootstrapping homomorphic encryption in
  less than a second,'' in \emph{Annual International Conference on the Theory
  and Applications of Cryptographic Techniques}.\hskip 1em plus 0.5em minus
  0.4em\relax Springer, 2015, pp. 617--640.

\bibitem{chillotti2016faster}
I.~Chillotti, N.~Gama, M.~Georgieva, and M.~Izabachene, ``Faster fully
  homomorphic encryption: Bootstrapping in less than 0.1 seconds,'' in
  \emph{International Conference on the Theory and Application of Cryptology
  and Information Security}.\hskip 1em plus 0.5em minus 0.4em\relax Springer,
  2016, pp. 3--33.

\bibitem{shih2017t}
M.-W. Shih, S.~Lee, T.~Kim, and M.~Peinado, ``T-sgx: Eradicating
  controlled-channel attacks against enclave programs,'' in \emph{Proceedings
  of the 2017 Annual Network and Distributed System Security Symposium (NDSS),
  San Diego, CA}, 2017.

\bibitem{chen2017detecting}
S.~Chen, X.~Zhang, M.~K. Reiter, and Y.~Zhang, ``Detecting privileged
  side-channel attacks in shielded execution with d{\'e}j{\'a} vu,'' in
  \emph{Proceedings of the 2017 ACM on Asia Conference on Computer and
  Communications Security}.\hskip 1em plus 0.5em minus 0.4em\relax ACM, 2017,
  pp. 7--18.

\bibitem{gotzfried2017cache}
J.~G{\"o}tzfried, M.~Eckert, S.~Schinzel, and T.~M{\"u}ller, ``Cache attacks on
  intel sgx,'' in \emph{Proceedings of the 10th European Workshop on Systems
  Security}.\hskip 1em plus 0.5em minus 0.4em\relax ACM, 2017, p.~2.

\bibitem{hahnel2017high}
M.~H{\"a}hnel, W.~Cui, and M.~Peinado, ``High-resolution side channels for
  untrusted operating systems,'' in \emph{2017 USENIX Annual Technical
  Conference (USENIX ATC 17)}, 2017, pp. 299--312.

\bibitem{gras2018translation}
B.~Gras, K.~Razavi, H.~Bos, and C.~Giuffrida, ``Translation leak-aside buffer:
  Defeating cache side-channel protections with {TLB} attacks,'' in \emph{27th
  $\{$USENIX$\}$ Security Symposium ($\{$USENIX$\}$ Security 18)}.\hskip 1em
  plus 0.5em minus 0.4em\relax {USENIX} Association, 2018.

\bibitem{moghimi2018memjam}
A.~Moghimi, T.~Eisenbarth, and B.~Sunar, ``Memjam: A false dependency attack
  against constant-time crypto implementations in sgx,'' in
  \emph{Cryptographers’ Track at the RSA Conference}.\hskip 1em plus 0.5em
  minus 0.4em\relax Springer, 2018, pp. 21--44.

\bibitem{yan2019attack}
M.~Yan, R.~Sprabery, B.~Gopireddy, C.~Fletcher, R.~Campbell, and J.~Torrellas,
  ``Attack directories, not caches: Side channel attacks in a non-inclusive
  world,'' in \emph{To appear in 2019 IEEE Symposium on Security and Privacy
  (SP)}.\hskip 1em plus 0.5em minus 0.4em\relax IEEE.

\bibitem{van2018malicious}
S.~van Schaik, C.~Giuffrida, H.~Bos, and K.~Razavi, ``Malicious management
  unit: Why stopping cache attacks in software is harder than you think,'' in
  \emph{27th {USENIX} Security Symposium ({USENIX} Security 18)}.\hskip 1em
  plus 0.5em minus 0.4em\relax Baltimore, MD: {USENIX} Association, 2018.

\bibitem{pessl2016drama}
P.~Pessl, D.~Gruss, C.~Maurice, M.~Schwarz, and S.~Mangard, ``Drama: Exploiting
  dram addressing for cross-cpu attacks.'' in \emph{USENIX Security Symposium},
  2016, pp. 565--581.

\bibitem{evtyushkin2016covert}
D.~Evtyushkin and D.~Ponomarev, ``Covert channels through random number
  generator: Mechanisms, capacity estimation and mitigations,'' in
  \emph{Proceedings of the 2016 ACM SIGSAC conference on computer and
  communications security}.\hskip 1em plus 0.5em minus 0.4em\relax ACM, 2016,
  pp. 843--857.

\bibitem{andrysco2015subnormal}
M.~Andrysco, D.~Kohlbrenner, K.~Mowery, R.~Jhala, S.~Lerner, and H.~Shacham,
  ``On subnormal floating point and abnormal timing,'' in \emph{IEEE Symposium
  on Security and Privacy (SP)}.\hskip 1em plus 0.5em minus 0.4em\relax IEEE,
  2015, pp. 623--639.

\bibitem{van2017sgx}
J.~Van~Bulck, F.~Piessens, and R.~Strackx, ``Sgx-step: A practical attack
  framework for precise enclave execution control,'' in \emph{Proceedings of
  the 2nd Workshop on System Software for Trusted Execution}.\hskip 1em plus
  0.5em minus 0.4em\relax ACM, 2017, p.~4.

\bibitem{lipp2018meltdown}
M.~Lipp, M.~Schwarz, D.~Gruss, T.~Prescher, W.~Haas, S.~Mangard, P.~Kocher,
  D.~Genkin, Y.~Yarom, and M.~Hamburg, ``Meltdown,'' \emph{arXiv preprint
  arXiv:1801.01207}, 2018.

\bibitem{kocher2018spectre}
P.~Kocher, D.~Genkin, D.~Gruss, W.~Haas, M.~Hamburg, M.~Lipp, S.~Mangard,
  T.~Prescher, M.~Schwarz, and Y.~Yarom, ``Spectre attacks: Exploiting
  speculative execution,'' \emph{arXiv preprint arXiv:1801.01203}, 2018.

\bibitem{chen2018sgxpectre}
G.~Chen, S.~Chen, Y.~Xiao, Y.~Zhang, Z.~Lin, and T.~H. Lai, ``Sgxpectre
  attacks: Leaking enclave secrets via speculative execution,'' \emph{arXiv
  preprint arXiv:1802.09085}, 2018.

\bibitem{bulck2018foreshadow}
\BIBentryALTinterwordspacing
J.~V. Bulck, M.~Minkin, O.~Weisse, D.~Genkin, B.~Kasikci, F.~Piessens,
  M.~Silberstein, T.~F. Wenisch, Y.~Yarom, and R.~Strackx, ``Foreshadow:
  Extracting the keys to the intel {SGX} kingdom with transient out-of-order
  execution,'' in \emph{27th {USENIX} Security Symposium ({USENIX} Security
  18)}.\hskip 1em plus 0.5em minus 0.4em\relax Baltimore, MD: {USENIX}
  Association, 2018, p. 991{\textendash}1008. [Online]. Available:
  \url{https://www.usenix.org/conference/usenixsecurity18/presentation/bulck}
\BIBentrySTDinterwordspacing

\bibitem{schuster2015vc3}
F.~Schuster, M.~Costa, C.~Fournet, C.~Gkantsidis, M.~Peinado, G.~Mainar-Ruiz,
  and M.~Russinovich, ``Vc3: Trustworthy data analytics in the cloud using
  sgx,'' in \emph{IEEE Symposium on Security and Privacy (SP)}.\hskip 1em plus
  0.5em minus 0.4em\relax IEEE, 2015, pp. 38--54.

\bibitem{zheng2017opaque}
W.~Zheng, A.~Dave, J.~G. Beekman, R.~A. Popa, J.~E. Gonzalez, and I.~Stoica,
  ``Opaque: An oblivious and encrypted distributed analytics platform.'' in
  \emph{NSDI}, 2017, pp. 283--298.

\bibitem{duan2017lightbox}
H.~Duan, X.~Yuan, and C.~Wang, ``Lightbox: Sgx-assisted secure network
  functions at near-native speed,'' \emph{arXiv preprint arXiv:1706.06261},
  2017.

\bibitem{kim2017enhancing}
S.~M. Kim, J.~Han, J.~Ha, T.~Kim, and D.~Han, ``Enhancing security and privacy
  of tor's ecosystem by using trusted execution environments.'' in \emph{NSDI},
  2017, pp. 145--161.

\bibitem{coughlin2017trusted}
M.~Coughlin, E.~Keller, and E.~Wustrow, ``Trusted click: Overcoming security
  issues of nfv in the cloud,'' in \emph{Proceedings of the ACM International
  Workshop on Security in Software Defined Networks \& Network Function
  Virtualization}.\hskip 1em plus 0.5em minus 0.4em\relax ACM, 2017, pp.
  31--36.

\bibitem{chen2016premix}
F.~Chen, M.~Dow, S.~Ding, Y.~Lu, X.~Jiang, H.~Tang, and S.~Wang, ``Premix:
  Privacy-preserving estimation of individual admixture,'' in \emph{AMIA Annual
  Symposium Proceedings}, vol. 2016.\hskip 1em plus 0.5em minus 0.4em\relax
  American Medical Informatics Association, 2016, p. 1747.

\bibitem{chen2016princess}
F.~Chen, S.~Wang, X.~Jiang, S.~Ding, Y.~Lu, J.~Kim, S.~C. Sahinalp, C.~Shimizu,
  J.~C. Burns, V.~J. Wright \emph{et~al.}, ``Princess: Privacy-protecting rare
  disease international network collaboration via encryption through software
  guard extensions,'' \emph{Bioinformatics}, vol.~33, no.~6, pp. 871--878,
  2016.

\bibitem{chen2017presage}
F.~Chen, C.~Wang, W.~Dai, X.~Jiang, N.~Mohammed, M.~M. Al~Aziz, M.~N. Sadat,
  C.~Sahinalp, K.~Lauter, and S.~Wang, ``Presage: Privacy-preserving genetic
  testing via software guard extension,'' \emph{BMC medical genomics}, vol.~10,
  no.~2, p.~48, 2017.

\bibitem{priebe2018enclavedb}
\BIBentryALTinterwordspacing
C.~Priebe, K.~Vaswani, and M.~Costa, ``Enclavedb: A secure database using
  sgx,'' in \emph{IEEE Symposium on Security and Privacy (SP)}, May 2018, pp.
  264--278. [Online]. Available:
  \url{doi.ieeecomputersociety.org/10.1109/SP.2018.00025}
\BIBentrySTDinterwordspacing

\bibitem{eskandarian2017oblivious}
S.~Eskandarian and M.~Zaharia, ``An oblivious general-purpose sql database for
  the cloud,'' \emph{arXiv preprint arXiv:1710.00458}, 2017.

\bibitem{gribov2017stealthdb}
A.~Gribov, D.~Vinayagamurthy, and S.~Gorbunov, ``Stealthdb: a scalable
  encrypted database with full sql query support,'' \emph{arXiv preprint
  arXiv:1711.02279}, 2017.

\bibitem{eskandarian2017oblidb}
S.~Eskandarian and M.~Zaharia, ``Oblidb: Oblivious query processing using
  hardware enclaves,'' \emph{arXiv preprint arXiv:1710.00458}, 2017.

\bibitem{gupta2016using}
D.~Gupta, B.~Mood, J.~Feigenbaum, K.~Butler, and P.~Traynor, ``Using intel
  software guard extensions for efficient two-party secure function
  evaluation,'' in \emph{International Conference on Financial Cryptography and
  Data Security}.\hskip 1em plus 0.5em minus 0.4em\relax Springer, 2016, pp.
  302--318.

\bibitem{kuccuk2016exploring}
K.~A. K{\"u}{\c{c}}{\"u}k, A.~Paverd, A.~Martin, N.~Asokan, A.~Simpson, and
  R.~Ankele, ``Exploring the use of intel sgx for secure many-party
  applications,'' in \emph{Proceedings of the 1st Workshop on System Software
  for Trusted Execution}.\hskip 1em plus 0.5em minus 0.4em\relax ACM, 2016,
  p.~5.

\bibitem{fisch2017iron}
B.~Fisch, D.~Vinayagamurthy, D.~Boneh, and S.~Gorbunov, ``Iron: functional
  encryption using intel sgx,'' in \emph{Proceedings of the 2017 ACM SIGSAC
  Conference on Computer and Communications Security}.\hskip 1em plus 0.5em
  minus 0.4em\relax ACM, 2017, pp. 765--782.

\bibitem{jiang2018securelr}
Y.~Jiang, J.~Hamer, C.~Wang, X.~Jiang, M.~Kim, Y.~Song, Y.~Xia, N.~Mohammed,
  M.~N. Sadat, and S.~Wang, ``Securelr: Secure logistic regression model via a
  hybrid cryptographic protocol,'' \emph{IEEE/ACM Transactions on Computational
  Biology and Bioinformatics}, 2018.

\bibitem{sadat2018secure}
M.~N. Sadat, X.~Jiang, M.~M. Al~Aziz, S.~Wang, and N.~Mohammed, ``Secure and
  efficient regression analysis using a hybrid cryptographic framework:
  Development and evaluation,'' \emph{JMIR medical informatics}, vol.~6, no.~1,
  2018.

\bibitem{sadat2018safety}
M.~N. Sadat, M.~M. Al~Aziz, N.~Mohammed, F.~Chen, X.~Jiang, and S.~Wang,
  ``Safety: Secure gwas in federated environment through a hybrid solution,''
  \emph{IEEE/ACM Transactions on Computational Biology and Bioinformatics},
  2018.

\bibitem{chenghong2017scotch}
C.~Wang, Y.~Jiang, N.~Mohammed, F.~Chen, X.~Jiang, M.~M. Al~Aziz, M.~N. Sadat,
  and S.~Wang, ``Scotch: Secure counting of encrypted genomic data using a
  hybrid approach,'' in \emph{AMIA Annual Symposium Proceedings}, vol.
  2017.\hskip 1em plus 0.5em minus 0.4em\relax American Medical Informatics
  Association, 2017, p. 1744.

\end{thebibliography}

\end{document}